\documentclass[12pt]{article}
\usepackage{amsmath,amsfonts,amssymb,bm}
\usepackage{graphicx}

\begin{document}

\title{An expression of excess work during transition between nonequilibrium steady states}

\author{Tatsuro Yuge
\\
\normalsize Department of Physics, Osaka University, 1-1, Toyonaka, Osaka, 560-0043, Japan
}

\maketitle

\begin{abstract}
Excess work is a non-diverging part of the work during transition between nonequilibrium steady states (NESSs). 
It is a central quantity in the steady state thermodynamics (SST), 
which is a candidate for nonequilibrium thermodynamics theory. 
We derive an expression of excess work during quasistatic transitions between NESSs 
by using the macroscopic linear response relation of NESS. 
This expression is a line integral of a vector potential in the space of control parameters. 
We show a relationship between the vector potential and the response function of NESS, 
and thus obtain a relationship between the SST and a macroscopic quantity. 
We also connect the macroscopic formulation to microscopic physics 
through a microscopic expression of the nonequilibrium response function, 
which gives a result consistent with the previous studies.  
\end{abstract}

\section{Introduction}
The second law of thermodynamics gives a fundamental limit to 
thermodynamic operations on systems in equilibrium states. 
One of its formulations provides the lower bound for the work performed on a system 
during a thermodynamics operation that induces a transition between equilibrium states; 
the lower bound is given by the change in the free energy and is achieved for quasistatic operations. 
To establish analogous thermodynamic theory for nonequilibrium steady state (NESS) 
is one of the challenging problems in physics. 
In recent attempts \cite{Esposito_etal07, KNST, EspositoBroeck, DeffnerLutz2010, DeffnerLutz2012, Takara_etal, SaitoTasaki,
SagawaHayakawa, Nakagawa, VerleyLacoste, Boksenbojm_etal, Mandal, MaesNetocny, Bertini_etal, Yuge_etal}, 
particularly investigated are relations to be satisfied 
by the work and entropy production (or heat) for transitions between NESSs. 

One of the candidates for nonequilibrium thermodynamics is the steady state thermodynamics (SST) 
proposed in \cite{OonoPaniconi}. 
A central idea of the SST is to use excess work (and excess heat), which is defined as follows. 
In nonequilibrium states, work is continuously supplied to the system, 
so that the total work $W_{\rm tot}$ during the transition between NESSs diverges. 
Therefore, for the construction of a meaningful thermodynamic theory for the transition, 
it is necessary to take a finite part out of $W_{\rm tot}$. 
To this end, the excess work $W_{\rm ex}$ is defined by subtracting from $W_{\rm tot}$ 
the integral of the steady work flow in the instantaneous NESS 
at each point of the operation \cite{OonoPaniconi,Landauer}. 

Studies along this idea have been developed in Refs.~\cite{KNST, SagawaHayakawa, SaitoTasaki, MaesNetocny, Bertini_etal, Yuge_etal, HatanoSasa, SasaTasaki}. 
We here concentrate our attention on quasistatic transitions. 
In the regime near to equilibrium, the work version of the results in Refs.~\cite{KNST, SaitoTasaki} 
states that the excess work $W_{\rm ex}$ for quasistatic transitions 
is given by the change in a certain scalar potential. 
In the regime far from equilibrium, by contrast, the work version of Refs.~\cite{SagawaHayakawa, Yuge_etal} 
states that in general it is not equal to the change in any scalar function but equal to a geometrical quantity; 
i.e., it is given by a line integral of a vector potential $\bm{A}^W$ in the operation parameter space. 
This suggests that $\bm{A}^W$ plays an important role in SST. 

Since these studies are based on the microscopic dynamics, 
the relationship with the microscopic state has been developed. 
In contrast, the relationship with macroscopic quantities has been less clear 
although there exist some studies on this issue \cite{Boksenbojm_etal,Mandal,Maes}. 
Particularly important is to clarify how the transport coefficients and response functions 
are treated in the framework of SST, 
because the transport coefficients are main quantities to characterize NESS 
and are accessible in experiments. 

In this paper, as a first step toward this issue, we derive an expression of $W_{\rm ex}$ for quasistatic transitions 
in terms of the linear response function of NESS. 
We show that $W_{\rm ex}$ is equal to the line integral of a vector potential $\bm{A}^W$ 
and clarify the relation between $\bm{A}^W$ and the response function. 
Since the derivation relies only on the macroscopic phenomenological equation (linear response relation), 
the result is universally valid (independent of microscopic detail).  
We also show that in the regime near to equilibrium $W_{\rm ex}$ is given by the change of a scalar potential 
thanks to the reciprocal relation. 
Furthermore, we connect the macroscopic theory for $\bm{A}^W$ to microscopic physics 
by using a microscopic expression of the response function, 
which is called response-correlation relation (RCR) \cite{ShimizuYuge,Yuge}. 
We obtain a microscopic expression of $\bm{A}^W$ that is consistent with 
the work version of the results in Refs.~\cite{SagawaHayakawa, Yuge_etal}.

\section{Setup}

\begin{figure}[t]
\begin{center}
\includegraphics[width=0.6\linewidth]{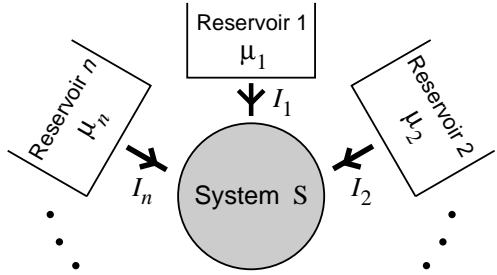}
\label{fig:illustration1}
\end{center}
\caption{
A schematic diagram of the setup. The system S is connected to $n$ reservoirs. 
The $i$th reservoir is characterized by the chemical potential $\mu_i$. 
The particle current $I_i$ flows from the $i$th reservoir into the system. 
}
\label{fig:schematic}
\end{figure}

We consider a system S that is in contact with multiple (say $n$) reservoirs.
A schematic diagram of the setup is shown in Fig.~\ref{fig:schematic}. 
The $i$th reservoir is in the equilibrium state characterized by the chemical potential $\mu_i$. 
We denote the set of the chemical potentials by $\mu$, i.e., $\mu=\{ \mu_i \}_{i=1}^n$.
The temperatures of all the reservoirs are set to the same value. 
We denote the particle current between S and the $i$th reservoir by $I_i$, 
where we take the sign of $I_i$ positive when it flows from the reservoir to S. 
More precisely, $I_i$ at time $t$ is defined as $I_i(t) = -dN_i(t)/dt$, 
where $N_i$ is the particle number in the $i$th reservoir. 
We assume that the reservoirs are sufficiently large so that they are not affected by the change in $N_i$ 
and remain in the equilibrium states on the time scale of interest.
We also assume that for a fixed $\mu$ a stable NESS is realized in S
uniquely and independently of initial states after a relaxation time.
We note that in the NESS $\sum_i \langle I_i \rangle_\mu^{\rm ss} = 0$ holds
due to the particle number conservation in the total system (S plus reservoirs), 
where $\langle I_i \rangle_\mu^{\rm ss}$ is the expectation value of the current $I_i$ 
in the NESS characterized by $\mu$. 

Examples of such a setup are seen in field-effect semiconductor devices. 
A typical one is the modulation-doped field-effect transistor (MODFET) \cite{Sze,Davies}. 
In this example, the system S is realized as the two-dimensional electron system, 
and the reservoirs are the electrodes (source, drain, and gate). 

\subsection{Transition between NESSs}

Suppose that at the initial time $t_0$ the system S is in a NESS characterized by $\mu$. 
The difference $\mu_i - \mu_j$ may be so large that the initial NESS is far from equilibrium. 
At $t_0+0$, we change the chemical potentials from $\mu$ to 
$\mu' = \{ \mu_i + \delta\mu_i \}_{i=1}^n$ with small constants $\{ \delta\mu_i \}_{i=1}^n$. 
Then the state of the system S varies in time for $t>t_0$, and 
after a sufficiently long time it settles to a new NESS characterized by $\mu'$. 
In this paper we investigate the work $W$ done on S during the transition between the NESSs. 

For this purpose, we here consider the expectation value of the current $I_i$ 
from the macroscopic viewpoint of the linear response relation. 
To the linear order in $\delta\mu$, we can express the expectation value $\langle I_i \rangle_{\mu'}^t$ at $t>t_0$ as 
\begin{align}
\langle I_i \rangle_{\mu'}^t 
= \langle I_i \rangle_\mu^{\rm ss} + \sum_{j=1}^n \Psi_{ij}(t-t_0) \delta\mu_j + O(\delta\mu^2), 
\label{linearResponse}
\end{align}
where $\Psi_{ij}(\tau) \equiv \int_0^\tau d\tau' \Phi_{ij}(\tau')$, 
and $\Phi_{ij}(\tau)$ is the linear response function of the NESS 
\cite{Marconi_etal, ChetriteGupta, Baiesi_etal, BaiesiMaes}, 
which satisfies the causality relation $\Phi_{ij}(\tau<0)=0$. 
We note that the linear response relation (\ref{linearResponse}) is a relation around a NESS (not equilibrium state) 
and is valid for a NESS even far from equilibrium if the NESS is stable to perturbations. 
Also we can express the expectation value of the current $I_i$ in the final NESS  (characterized by $\mu'$)
by the long-time limit of Eq.~(\ref{linearResponse}): 
\begin{align}
\langle I_i \rangle_{\mu'}^{\rm ss} 
= \langle I_i \rangle_\mu^{\rm ss} + \sum_j \tilde{\Phi}_{ij} \delta\mu_j + O(\delta\mu^2), 
\label{steadyCurrentAfter}
\end{align}
where $\tilde{\Phi}_{ij} \equiv \lim_{\tau\to\infty}\Psi_{ij}(\tau)$ 
is the transport coefficient (differential conductivity) of the initial NESS (characterized by $\mu$).
We again note that the limit exists if the NESS is stable.

\section{Excess work}

\subsection{General case}

In nonequilibrium states, work $W$ is continuously supplied to the system S from the reservoirs, 
accompanied by the particle current to S. 
We can use the equilibrium thermodynamics to estimate the work done by the reservoirs 
since they are in the equilibrium states; 
When the particle number in the $i$th reservoir increases by $\Delta N_i$, 
the work done by the $i$th reservoir is given by $W_i = - \mu'_i \Delta N_i$
Therefore, the unit-time work by the $i$th reservoir is $J_i^W = \mu'_i I_i$. 
From Eq.~(\ref{linearResponse}), we obtain the average work flow 
$\langle J^W \rangle_{\mu'}^t = \sum_i \langle J_i^W \rangle_{\mu'}^t$ at time $t>t_0$ as 
\begin{align}
\langle J^W \rangle_{\mu'}^t &= \sum_i \langle I_i \rangle_{\mu'}^t \bigl( \mu_i + \delta\mu_i \bigr) 
\nonumber\\
&\simeq \sum_i \langle I_i \rangle_\mu^{\rm ss} \bigl( \mu_i + \delta\mu_i \bigr) 
+ \sum_{ij} \mu_i \Psi_{ij}(t-t_0) \delta\mu_j, 
\label{workFlow_t}
\end{align}
where we neglected the $O(\delta\mu^2)$ terms in the second line. 
The total work $W_{\rm tot}$ during the transition between the NESSs 
is given by the time-integral of Eq.~(\ref{workFlow_t}). 
However, $W_{\rm tot}$ is a diverging quantity because the system S remains to be supplied with the work 
from the reservoirs after it reaches the final NESS. 

To extract a finite quantity intrinsic to the transition, we employ the idea of the SST \cite{OonoPaniconi}; 
we subtract from $W_{\rm tot}$ the contribution 
of the steady work flow $\langle J^W \rangle_{\mu'}^{\rm ss}$ in the final NESS: 
\begin{align}
W_{\rm ex} 
\equiv \int_{t_0}^\infty dt \Bigl( \langle J^W \rangle_{\mu'}^t - \langle J^W \rangle_{\mu'}^{\rm ss} \Bigr). 
\label{excessWork}
\end{align}
We refer to this quantity as the excess work. 
We note that $W_{\rm ex}$ is related to the excess heat $Q_{\rm ex}$ as $W_{\rm ex} + Q_{\rm ex} = \Delta U$, 
where $\Delta U$ is the change in the energy of S between the NESSs, 
due to the energy conservation in the transition between the NESSs and the energy balance in the steady states.  
Our definition (\ref{excessWork}) of $W_{\rm ex}$ is consistent with the definition of $Q_{\rm ex}$ 
in Refs.~\cite{KNST,SaitoTasaki,SagawaHayakawa,Yuge_etal}. 
We also note that the steady flow $\langle J^W \rangle_{\mu'}^{\rm ss}$ is equal to 
the long-time limit of Eq.~(\ref{workFlow_t}): 
\begin{align}
\langle J^W \rangle_{\mu'}^{\rm ss} 
&= \sum_i \langle I_i \rangle_\mu^{\rm ss} \bigl( \mu_i + \delta\mu_i \bigr) 
+ \sum_{ij} \mu_i \tilde{\Phi}_{ij} \delta\mu_j. 
\label{workFlow_ss}
\end{align}

By substituting  Eqs.~(\ref{workFlow_t}) and (\ref{workFlow_ss}) into Eq.~(\ref{excessWork}), we obtain 
\begin{align}
W_{\rm ex} = \sum_j A_j^W \delta\mu_j, 
\label{geometricalExcessWork}
\end{align}
where the $j$th component $A_j^W$ of the vector potential $\bm{A}^W$ is given by 
\begin{align}
A_j^W = \sum_i \mu_i \int_{t_0}^\infty dt \Bigl[ \Psi_{ij}(t-t_0) - \tilde{\Phi}_{ij} \Bigr]. 
\label{vectorPotentialResponse}
\end{align}
Equation~(\ref{geometricalExcessWork}) indicates that 
the excess work during quasistatic transitions between NESSs 
is not given by the difference of some scalar function $F$ but given by the geometrical quantity 
unless $A_j^W$ is equal to the $\mu_j$-derivative of $F$ for all $j$. 
This is consistent with the results in \cite{SagawaHayakawa,Yuge_etal}. 
Equation~(\ref{vectorPotentialResponse}) relates the nonequilibrium linear response function $\Phi$ 
with the vector potential $\bm{A}^W$ in the expression (\ref{geometricalExcessWork}). 
Therefore $\bm{A}^W$ can be experimentally determined in principle, because $\Phi$ is measurable. 

The sufficient condition for $A_j^W = \partial_j F$ for all $j$ is that 
\begin{align}
\partial_i A_j^W = \partial_j A_i^W 
\label{MaxwellRelation}
\end{align}
holds for all $i,j$, where $\partial_j$ is the abbreviation of $\partial/\partial \mu_j$.

\subsection{Weakly nonequilibrium case}
In the regime near to equilibrium (linear response regime), 
we can use the response function $\Phi^{\rm eq}$ of the equilibrium state in Eq.~(\ref{vectorPotentialResponse}). 
Then Eq.~(\ref{MaxwellRelation}) is valid because $\Phi^{\rm eq}$ is independent of $\mu_i$ 
and the reciprocal relation $\Phi^{\rm eq}_{ij} = \Phi^{\rm eq}_{ji}$ holds. 
Therefore the extension of the Clausius equality is possible in this regime,
which is consistent with the results in Refs.~\cite{KNST,SaitoTasaki}.

\section{Connection to microscopic physics}\label{sec:micro}

Up to here our formulation is closed on the macroscopic level. 
Now we connect it to microscopic physics. 
In this paper we assume that the microscopic dynamics of the system S is governed 
by the quantum master equation (QME) \cite{BreuerPetruccione}: 
\begin{align}
\frac{\partial \hat{\rho}}{\partial t} = \mathcal{K}\hat{\rho}. 
\label{QME}
\end{align}
Here, $\hat{\rho}$ is the density matrix of S, and the generator $\mathcal{K}$ is written as 
$\mathcal{K} = [\hat{H}, ~]/i\hbar + \sum_j \mathcal{L}_j$, 
where $\hat{H}$ is the Hamiltonian of S 
and $\mathcal{L}_j$ is the dissipator induced by the interaction with the $j$th reservoir. 
As in the previous sections, we assume that there exists a unique steady state in the QME.
The steady state density matrix $\hat{\rho}_{\rm ss}$ satisfies $\mathcal{K}\hat{\rho}_{\rm ss}=0$.

First, we consider the connection to microscopic physics through the response function $\Phi$ of NESS. 
For this purpose we employ the response-correlation relation (RCR) \cite{ShimizuYuge,Yuge}, 
which is a microscopic expression of $\Phi$. 
In the framework of the QME and for the response of the current $I_i$ from the $i$th reservoir into S, 
the RCR reads 
\begin{align}
\Phi_{ij}(\tau) 
= {\rm Tr} \bigl[ (\partial_j \hat{I}_i) \hat{\rho}_{\rm ss} \bigr] \delta(\tau) 
+ {\rm Tr} \bigl[ \hat{I}_i e^{\mathcal{K}\tau} (\partial_j \mathcal{K}) \hat{\rho}_{\rm ss} \bigr], 
\label{RCR}
\end{align}
where ${\rm Tr}$ is the trace over S and $\hat{I}_i \equiv \mathcal{L}_i^\dag \hat{N}$ 
with $\hat{N}$ being the particle number operator in S.
See \ref{appendix:RCR} for the derivation of Eq.~(\ref{RCR}). 
Note that $\hat{I}_i$ can be regarded as the particle current operator from the $i$th reservoir into S 
because it satisfies the continuity equation: 
$(\partial/\partial t) {\rm Tr} [ \hat{N} \hat{\rho}(t)] = \sum_i {\rm Tr} [\hat{I}_i \hat{\rho}(t)]$.
Substituting Eq.~(\ref{RCR}) into Eq.~(\ref{vectorPotentialResponse}), we obtain 
\begin{align}
A_j^W = \sum_i \mu_i \int_{t_0}^\infty dt \biggl\{ & \int_{t_0}^t dt' 
{\rm Tr} \bigl[ \hat{I}_i e^{\mathcal{K}(t-t')} (\partial_j \mathcal{K}) \hat{\rho}_{\rm ss} \bigr] 
+ {\rm Tr} \bigl[ \hat{I}_i \mathcal{R} (\partial_j \mathcal{K}) \hat{\rho}_{\rm ss} \bigr] \biggr\}. 
\label{vectorPotentialRCR}
\end{align}
Note that the contribution from the first term in Eq.~(\ref{RCR}) vanishes. 
Here we defined 
\begin{align}
\mathcal{R} \equiv -\lim_{T\to\infty}\int_{t_0}^T dt' e^{\mathcal{K}(T-t')} \mathcal{Q}_0, 
\label{R}
\end{align}
and $\mathcal{Q}_0 = 1- \mathcal{P}_0$, 
where the projection superoperator $\mathcal{P}_0$ is defined such that 
$\mathcal{P}_0 \hat{X}= \hat{\rho}_{\rm ss} {\rm Tr}\hat{X}$ holds for any linear operator $\hat{X}$. 
See \ref{appendix:inverse} for the fact that $\mathcal{R}$ is a well-defined superopertor. 
To rewrite Eq.~(\ref{vectorPotentialRCR}) further, we note the following relation:
\begin{align}
\frac{d}{dt'}{\rm Tr}\bigl[ \hat{I}_i e^{\mathcal{K}(t-t')} \mathcal{R} (\partial_j \mathcal{K}) \hat{\rho}_{\rm ss} \bigr]
&= -{\rm Tr}\bigl[ \hat{I}_i e^{\mathcal{K}(t-t')} \mathcal{K} \mathcal{R} (\partial_j \mathcal{K}) \hat{\rho}_{\rm ss} \bigr]
\nonumber\\
&= -{\rm Tr}\bigl[ \hat{I}_i e^{\mathcal{K}(t-t')} \mathcal{Q}_0 (\partial_j \mathcal{K}) \hat{\rho}_{\rm ss} \bigr]
\nonumber\\
&= -{\rm Tr}\bigl[ \hat{I}_i e^{\mathcal{K}(t-t')} (\partial_j \mathcal{K}) \hat{\rho}_{\rm ss} \bigr]
+ {\rm Tr}\bigl[ \hat{I}_i \hat{\rho}_{\rm ss} \bigr] {\rm Tr}\bigl[(\partial_j \mathcal{K}) \hat{\rho}_{\rm ss} \bigr]
\nonumber\\
&= -{\rm Tr}\bigl[ \hat{I}_i e^{\mathcal{K}(t-t')} (\partial_j \mathcal{K}) \hat{\rho}_{\rm ss} \bigr].
\label{integrand_vec}
\end{align}
In the third line we used 
\begin{align}
\mathcal{R}\mathcal{K} = \mathcal{K}\mathcal{R} = \mathcal{Q}_0. 
\label{RK}
\end{align}
See \ref{appendix:inverse} for the derivation of Eq.~(\ref{RK}).
In the last line of Eq.~(\ref{integrand_vec}) we used 
${\rm Tr}\bigl[(\partial_j \mathcal{K}) \hat{\rho}_{\rm ss} \bigr]
= \partial_j  {\rm Tr}\bigl[\mathcal{K} \hat{\rho}_{\rm ss} \bigr] 
- {\rm Tr}\bigl[\mathcal{K} \partial_j \hat{\rho}_{\rm ss} \bigr] = 0$. 
This follows from $\mathcal{K} \hat{\rho}_{\rm ss}=0$ and ${\rm Tr}\bigl[\mathcal{K} \hat{X}]=0$ for any $\hat{X}$
(trace-preserving property of the QME).
Integrating Eq.~(\ref{integrand_vec}), we can rewrite the first term 
on the right hand side of Eq.~(\ref{vectorPotentialRCR}) as 
\begin{align}
\int_{t_0}^t dt' {\rm Tr} \bigl[ \hat{I}_i e^{\mathcal{K}(t-t')} (\partial_j \mathcal{K}) \hat{\rho}_{\rm ss} \bigr] 
= {\rm Tr} \bigl[ \hat{I}_i e^{\mathcal{K}(t-t_0)} \mathcal{R} (\partial_j \mathcal{K}) \hat{\rho}_{\rm ss} \bigr] 
-{\rm Tr} \bigl[ \hat{I}_i \mathcal{R} (\partial_j \mathcal{K}) \hat{\rho}_{\rm ss} \bigr]. 
\nonumber
\end{align}
The second term on the right hand side of this equation cancels out 
the second term on the right hand side of Eq.~(\ref{vectorPotentialRCR}).
We thus rewrite Eq.~(\ref{vectorPotentialRCR}) as
\begin{align}
A_j^W = - \sum_i \mu_i {\rm Tr} \left[ \hat{I}_i \mathcal{R}^2 
(\partial_j \mathcal{K}) \hat{\rho}_{\rm ss} \right],
\label{vectorPotentialRR}
\end{align}
where we used $\int_{t_0}^\infty dt e^{\mathcal{K}(t-t_0)}\mathcal{Q}_0 = -\mathcal{R}$.
This is a microscopic expression of the vector potential $\bm{A}^W$. 

Next, we investigate the consistency of Eq.~(\ref{vectorPotentialRR}) 
with the results in Refs.~\cite{SagawaHayakawa,Yuge_etal}. 
In a manner almost the same as those in Refs.~\cite{SagawaHayakawa,Yuge_etal}, 
we can derive another microscopic expression of $\bm{A}^W$ 
without relying on Eq.~(\ref{vectorPotentialResponse}): 
\begin{align}
A_j^W = -{\rm Tr} \left( \hat{\ell}^{\prime\dag}_0 \partial_j \hat{\rho}_{\rm ss} \right), 
\label{vectorPotentialQME}
\end{align}
where $\hat{\ell}^\prime_0 \equiv \partial \hat{\ell}^\chi_0/\partial (i\chi) |_{\chi=0}$. 
Here, $\chi$ is the counting field in the full counting statistics (FCS) of the work $W$ from the reservoirs, 
and $\hat{\ell}^\chi_0$ is the left eigenvector of $\mathcal{K}^\chi$ 
corresponding to the eigenvalue $\lambda^\chi_0$ that has the maximum real part. 
$\mathcal{K}^\chi=[\hat{H},~]/i\hbar + \sum_j \mathcal{L}_j^\chi$ is the $\chi$-modified generator, 
which is introduced for the FCS in the framework of the QME \cite{EspositoHarbolaMukamel_RMP}. 
See \ref{appendix:vectorPotentialQME} for the details and derivation. 
We note that $\hat{\ell}^{\chi=0}_0=\hat{1}$ (identity operator), $\lambda^{\chi=0}_0 = 0$, 
and $\partial \lambda^\chi_0 / \partial (i\chi) |_{\chi=0} = \langle J^W \rangle^{\rm ss}_\mu$. 
In the following we rewrite Eq.~(\ref{vectorPotentialQME}) 
to show its equivalence to Eq.~(\ref{vectorPotentialRR}). 

We first rewrite $\hat{\ell}^{\prime\dag}_0$ in Eq.~(\ref{vectorPotentialQME}). 
By differentiating the left eigenvalue equation 
$(\mathcal{K}^\chi)^\dag \hat{\ell}^\chi_0 = (\lambda^\chi_0)^* \hat{\ell}^\chi_0$ with respect to $i\chi$ 
and setting $\chi=0$, we obtain 
\begin{align}
\mathcal{K}^\dag \hat{\ell}^\prime_0 
= - (\mathcal{K}')^\dag \hat{1} - \langle J^W \rangle^{\rm ss}_\mu \hat{1}. 
\label{KDag_ellPrime}
\end{align}
Here $\mathcal{K}' = \partial \mathcal{K}^\chi / \partial (i\chi) |_{\chi=0}$ 
and the adjoint $\mathcal{O}^\dag$ of a superoperator $\mathcal{O}$ is defined 
by ${\rm Tr} [(\mathcal{O}^\dag \hat{X}_1)^\dag \hat{X}_2] = {\rm Tr} [ \hat{X}_1^\dag \mathcal{O} \hat{X}_2]$ 
for any pair $(\hat{X}_1, \hat{X}_2)$ of linear operators.
By operating on the both sides of Eq.~(\ref{KDag_ellPrime}) with $\mathcal{R}^\dag$, we obtain 
\begin{align}
\hat{\ell}^\prime_0 = - \mathcal{R}^\dag (\mathcal{K}')^\dag \hat{1} + c \hat{1}, 
\label{ellPrime}
\end{align}
where $c = - \langle J^W \rangle^{\rm ss}_\mu + {\rm Tr} [\hat{\rho}_{\rm ss} \hat{\ell}^\prime_0]$ 
and we used Eq.~(\ref{RK}).
Substituting Eq.~(\ref{ellPrime}) into Eq.~(\ref{vectorPotentialQME}) 
and using ${\rm Tr} [\partial_j \hat{\rho}_{\rm ss}] = \partial_j {\rm Tr} [\hat{\rho}_{\rm ss}] = 0$, 
we have $A_j^W = {\rm Tr} \bigl[ \bigl( (\mathcal{K}')^\dag \hat{1} \bigr)^\dag 
\mathcal{R} \partial_j \hat{\rho}_{\rm ss} \bigr]$. 
Furthermore we can show 
$(\mathcal{K}')^\dag \hat{1} = \sum_i \mu_i \mathcal{L}_i^\dag \hat{N} = \sum_i \mu_i \hat{I}_i$.
With this equation we have 
\begin{align}
A_j^W = \sum_i \mu_i {\rm Tr} [ \hat{I}_i \mathcal{R} \partial_j \hat{\rho}_{\rm ss} ].
\label{vectorPotentialQME_middle}
\end{align}

We next rewrite $\partial_j \hat{\rho}_{\rm ss}$. 
By differentiating the steady-state equation $\mathcal{K} \hat{\rho}_{\rm ss} = 0$ with respect to $\mu_j$ 
and operating on it with $\mathcal{R}$, we have 
$\partial_j \hat{\rho}_{\rm ss} = - \mathcal{R} (\partial_j \mathcal{K}) \hat{\rho}_{\rm ss}$. 
Substituting this equation into Eq.~(\ref{vectorPotentialQME_middle}) we obtain 
\begin{align}
A_j^W = - \sum_i \mu_i {\rm Tr} \left[ \hat{I}_i \mathcal{R}^2 
(\partial_j \mathcal{K}) \hat{\rho}_{\rm ss} \right].
\nonumber
\end{align}
This is the same as Eq.~(\ref{vectorPotentialRR}).
We thus show that Eq.~(\ref{vectorPotentialRR}) [and therefore Eq.~(\ref{vectorPotentialResponse})] 
is consistent with the results in Refs.~\cite{SagawaHayakawa,Yuge_etal}.

\section{Concluding remarks}

We have derived an expression of the excess work for quasistatic transitions between NESSs 
in particle transport systems on the basis of the linear response relation. 
We have related the vector potential $\bm{A}^W$ in the expression with the response function. 
We note that it is possible to extend our formulation to situations 
where other control parameters for transition between NESS are varied. 
In particular, we can obtain a similar result in heat conducting systems, 
where the temperatures of heat reservoirs are changed. 
We finally make remarks. 

First, the relationship between the excess work and the response function suggests 
that the response functions can be calculated in the framework of the SST. 
We expect that this expression becomes a first step for understanding of 
how transport phenomena are treated in the SST.  

Second, as is mentioned below Eq.~(\ref{excessWork}), 
our definition of the excess work is consistent with the definition of the excess heat 
in Refs.~\cite{KNST,SaitoTasaki,SagawaHayakawa,Yuge_etal}. 
However the definition of the excess work and heat is not unique; 
e.g., there are Hatano-Sasa type \cite{HatanoSasa,DeffnerLutz2012,Mandal} 
and Maes-Neto\v{c}n\'{y} type \cite{MaesNetocny} approaches. 
Recently, Ref.~\cite{Mandal} gave evidence that the Hatano-Sasa approach is appropriate for the definition.
Since the Hatano-Sasa approach relies on microscopic information (e.g., steady-state distribution and transition rate),
the connection to macroscopic quantities is not clear.
It is therefore important to investigate the definition from the viewpoint of response function as a future work.

Third, in recent years the nonequilibrium response function is a hot topic in the statistical physics 
\cite{Marconi_etal, ChetriteGupta, Baiesi_etal, BaiesiMaes}. 
One of the points in recent works is decomposition of the response function \cite{Baiesi_etal, BaiesiMaes}. 
We expect that the application of these results to the expression of the excess work would lead to 
a further decomposition of the work that is appropriate for the construction of the SST.

\section*{Acknowledgments}
The author thanks K. Akiba and M. Yamaguchi for helpful discussions. 
This work was supported by a JSPS Research Fellowship for Young Scientists (No. 24-1112), 
KAKENHI (No. 26287087), 
and ImPACT Program of Council for Science, Technology and Innovation (Cabinet Office, Government of Japan).

\appendix

\section{Linear response function of NESS in quantum master equation approach}\label{appendix:RCR}

Here we derive Eq.~(\ref{RCR}), the RCR in QME. 
We consider the QME~(\ref{QME}) that depends on multiple parameters $\bm{\alpha}=(\alpha_1,\alpha_2,\alpha_3,...)$
like chemical potentials; i.e., we assume that the generator of the QME depends on these parameters: 
$\mathcal{K}=\mathcal{K}(\bm{\alpha})$. 

Suppose that at time $t \le t_0$ the system S is in the NESS with $\bm{\alpha}=\bm{\alpha}^0$; 
i.e., $\hat{\rho}(t) = \hat{\rho}_{\rm ss}^0$ for $t \le t_0$, 
where $\hat{\rho}_{\rm ss}^0$ satisfies $\mathcal{K}^0 \hat{\rho}_{\rm ss}^0=0$ 
with $\mathcal{K}^0 \equiv \mathcal{K}(\bm{\alpha}^0)$. 
For $t>t_0$, we weakly modulate the parameters in time: $\alpha_l(t) = \alpha_l^0 + f_l(t)$ 
($l=1,2,3,...$), where $\max_t |f_l(t)|$ is much smaller than a typical value of $\alpha_l$.
Then we can expand the generator $\mathcal{K}$ around $\bm{\alpha}^0$ in terms of $f$: 
\begin{align}
\mathcal{K}\bigl(\bm{\alpha}(t)\bigr) \simeq \mathcal{K}^0 + \sum_l f_l(t) \partial_l \mathcal{K}^0, 
\label{expandK}
\end{align} 
where 
$\partial_l \mathcal{K}^0 \equiv \partial \mathcal{K}(\bm{\alpha})/\partial \alpha_l |_{\bm{\alpha}=\bm{\alpha}^0}$. 

To solve the QME~(\ref{QME}) with the weakly time-dependent $\bm{\alpha}$ and 
the initial condition $\hat{\rho}(t_0) = \hat{\rho}_{\rm ss}^0$, 
we transform the QME into an ``interaction picture''. 
That is, we introduce $\breve{\rho}(t) = e^{-\mathcal{K}_0(t-t_0)} \hat{\rho}(t)$. 
Then, from the QME~(\ref{QME}), we have the equation of motion for $\breve{\rho}$ as 
\begin{align}
\frac{\partial \breve{\rho}(t)}{\partial t} 
= \sum_l f_l(t) e^{-\mathcal{K}_0(t-t_0)} (\partial_l \mathcal{K}^0) e^{\mathcal{K}_0(t-t_0)} \breve{\rho}(t), 
\end{align}
where we used Eq.~(\ref{expandK}). 
By integrating this equation form $t_0$ to $t$ with $\breve{\rho}(t_0) = \hat{\rho}_{\rm ss}^0$, 
we obtain 
\begin{align}
\breve{\rho}(t) &= \hat{\rho}_{\rm ss}^0 
+ \sum_l \int_{t_0}^t dt' f_l(t') e^{-\mathcal{K}_0(t'-t_0)} (\partial_l \mathcal{K}^0) 
e^{\mathcal{K}_0(t'-t_0)} \breve{\rho}(t') 
\nonumber\\
&\simeq \hat{\rho}_{\rm ss}^0 
+ \sum_l \int_{t_0}^t dt' f_l(t') e^{-\mathcal{K}_0(t'-t_0)} (\partial_l \mathcal{K}^0) \hat{\rho}_{\rm ss}^0. 
\end{align}
In going from the first line to the second, we approximately replaced $\breve{\rho}(t')$ in the integral 
with $\hat{\rho}_{\rm ss}^0$. 
This approximation corresponds to the first-order time-dependent perturbation theory in quantum mechanics. 
Going back to the Schr\"odinger picture, we have 
\begin{align}
\hat{\rho}(t) &= \hat{\rho}_{\rm ss}^0 
+ \sum_l \int_{t_0}^t dt' f_l(t') e^{\mathcal{K}_0(t-t')} (\partial_l \mathcal{K}^0) \hat{\rho}_{\rm ss}^0. 
\end{align}
We thus obtain the time dependence of the expectation value of a quantity $\hat{X}$ 
that is independent of $\bm{\alpha}$: 
\begin{align}
\langle X \rangle^t_{\bm{\alpha}(t)} &= {\rm Tr} \big[ \hat{X} \hat{\rho}(t) \bigr] 
\nonumber\\
&= \langle X \rangle^{\rm ss}_{\bm{\alpha}_0}
+ \sum_l \int_{t_0}^t dt' {\rm Tr} \Big[ \hat{X} e^{\mathcal{K}^0(t-t')} (\partial_l \mathcal{K}^0)
\hat{\rho}_{\rm ss}^0 \Bigr] f_l(t') 
\label{linearResponseQME1}
\\
&= \langle X \rangle^{\rm ss}_{\bm{\alpha}_0}
+ \sum_l \int_{t_0}^t dt' {\rm Tr} \Big[ 
\Bigl\{ (\partial_l \mathcal{K}^0)^\dag e^{\mathcal{K}^{0\dag}(t-t')} \hat{X} \Bigr\} 
\hat{\rho}_{\rm ss}^0 \Bigr] f_l(t'). 
\label{linearResponseQME2}
\end{align}
Equations (\ref{linearResponseQME1}) and (\ref{linearResponseQME2}) give the RCR in the QME.
We note that Eq.~(\ref{linearResponseQME2}) reduces to the Kubo formula 
if $\hat{\rho}_{\rm ss}^0$ is an equilibrium state 
(i.e., when we consider the response of an equilibrium state) \cite{ShimizuYuge}.

Now we consider the current $\hat{I}_i = \mathcal{L}_i^\dag \hat{N}$ 
from the $i$th reservoir into the system S. 
We note that $\hat{I}_i$ is dependent on $\bm{\alpha}$ because so is $\mathcal{L}_i$. 
Therefore we have the average current at time $t$ as 
\begin{align}
&\langle \hat{I}_i \rangle^t_{\bm{\alpha}(t)} - \langle \hat{I}_i \rangle^{\rm ss}_{\bm{\alpha}_0} 
\nonumber\\
&= {\rm Tr} \big[ \hat{I}_i\bigl(\bm{\alpha}(t)\bigr) \hat{\rho}(t) \bigr] 
- {\rm Tr} \big[ \hat{I}_i\bigl(\bm{\alpha}_0\bigr) \hat{\rho}^0_{\rm ss} \bigr] 
\nonumber\\
&= \sum_l {\rm Tr} \big[ (\partial_l \hat{I}_i^0) \hat{\rho}^0_{\rm ss} \bigr] f_l(t) 
+ {\rm Tr} \big[ \hat{I}_i^0 \hat{\rho}(t) \bigr] - {\rm Tr} \big[ \hat{I}_i^0 \hat{\rho}^0_{\rm ss} \bigr] 
\nonumber\\
&= \sum_l {\rm Tr} \big[ (\partial_l \hat{I}_i^0) \hat{\rho}^0_{\rm ss} \bigr] f_l(t) 
+ \sum_l \int_{t_0}^t dt' {\rm Tr} \Big[ \hat{I}_i^0 e^{\mathcal{K}^0(t-t')} (\partial_l \mathcal{K}^0)
\hat{\rho}_{\rm ss}^0 \Bigr] f_l(t'), 
\end{align}
where $\hat{I}_i^0 \equiv \hat{I}_i(\bm{\alpha}_0)$ and 
$\partial_l \hat{I}_i^0 \equiv \partial \hat{I}_i(\bm{\alpha})/\partial \alpha_l |_{\bm{\alpha}=\bm{\alpha}_0}$. 
We used Eq.~(\ref{linearResponseQME1}) in the third line. 
Finally, by performing the functional differentiation with respect to $f_j(t')$, we obtain 
\begin{align}
\Phi_{ij}(t-t') &= \frac{\delta \langle \hat{I}_i \rangle^t_{\bm{\alpha}(t)}}{\delta f_j(t')}
\nonumber\\
&= {\rm Tr} \bigl[ (\partial_j \hat{I}_i^0) \hat{\rho}_{\rm ss}^0 \bigr] \delta(t-t') 
+ {\rm Tr} \bigl[ \hat{I}_i^0 e^{\mathcal{K}(t-t')} (\partial_j \mathcal{K}) \hat{\rho}_{\rm ss}^0 \bigr]. 
\nonumber
\end{align}
This is equivalent to Eq.~(\ref{RCR}).

\section{Inverse-like superoperator in quantum master equation approach}\label{appendix:inverse}

First we show that $\mathcal{R}$ in Eq.~(\ref{R}) is well defined. 
To this end, we denote the eigenvalue and corresponding left and right eigenvectors of $\mathcal{K}$ 
as $\lambda_m$, $\hat{\ell}_m$, and $\hat{r}_m$. 
We assign the steady state of $\mathcal{K}$ to the index $m=0$; 
i.e., $\lambda_0 = 0$, $\hat{\ell}_0=\hat{1}$, and $\hat{r}_0 = \hat{\rho}_{\rm ss}$. 
By the assumption of the unique existence of the stable steady state, ${\rm Re}\lambda_m < 0$ for $m \neq 0$.
Then, for any linear operator $\hat{X}$, we obtain the following equation: 
\begin{align}
\mathcal{R} \hat{X} &= - \int_0^\infty dt e^{\mathcal{K}t} \mathcal{Q}_0 \hat{X}
\nonumber\\
&= - \int_0^\infty dt e^{\mathcal{K}t} \sum_{m\neq 0} {\rm Tr}[\hat{\ell}_m^\dag \hat{X}] \hat{r}_m
\nonumber\\
&= - \int_0^\infty dt \sum_{m\neq 0} e^{\lambda_m t} {\rm Tr}[\hat{\ell}_m^\dag \hat{X}] \hat{r}_m
\nonumber\\
&= \sum_{m\neq 0}  \frac{{\rm Tr}[\hat{\ell}_m^\dag \hat{X}]}{\lambda_m} \hat{r}_m.
\end{align}
Since this is not diverging, $\mathcal{R}$ is well defined. 

We here show that Eq.~(\ref{RK}) holds for the generator $\mathcal{K}$ of the QME. 
We first note that $\mathcal{R} \mathcal{K} = \mathcal{K} \mathcal{R}$ follows from 
$\mathcal{Q}_0 \mathcal{K} = \mathcal{K} \mathcal{Q}_0= \mathcal{K}$, which we can derive from the fact that 
\begin{align}
\mathcal{P}_0 \mathcal{K} \hat{X} &= \hat{\rho}_{\rm ss} {\rm Tr} [ \mathcal{K} \hat{X} ] = 0, 
\label{PKX}
\\
\mathcal{K} \mathcal{P}_0 \hat{X} &= \mathcal{K} \hat{\rho}_{\rm ss} {\rm Tr} \hat{X} = 0, 
\label{KPX}
\end{align}
hold for any linear operator $\hat{X}$. 
Equation~(\ref{PKX}) follows from ${\rm Tr} [ \mathcal{K} \hat{X} ] = 0$ (trace-preserving property of QME), 
and Eq.~(\ref{KPX}) from $\mathcal{K} \hat{\rho}_{\rm ss}=0$ (steady-state equation). 
Then we can show Eq.~(\ref{RK}) as follows: 
\begin{align}
\mathcal{R} \mathcal{K} = \mathcal{K} \mathcal{R} 
&= \lim_{T\to\infty} \int_{t_0}^T dt' \frac{d}{dt'}e^{\mathcal{K}(T-t')} \mathcal{Q}_0
\nonumber\\
&= \Bigl( 1 - \lim_{T\to\infty} e^{\mathcal{K}(T-t_0)} \Bigr) \mathcal{Q}_0
\nonumber\\
&= ( 1 - \mathcal{P}_0 ) \mathcal{Q}_0. 
\nonumber\\
&= \mathcal{Q}_0. 
\label{RK_a}
\end{align}
Here the third line follows from the convergence theorem of the Markov process, 
which we can derive from the fact that for any linear operator $\hat{X}$ the following equation holds: 
\begin{align}
\lim_{T\to\infty} e^{\mathcal{K}(T-t_0)} \hat{X} 
&= \lim_{T\to\infty} e^{\mathcal{K}(T-t_0)} \sum_m {\rm Tr}[\hat{\ell}_m^\dag \hat{X}] \hat{r}_m
\nonumber\\
&= \sum_m {\rm Tr}[\hat{\ell}_m^\dag \hat{X}] \hat{r}_m \lim_{T\to\infty} e^{\lambda_m(T-t_0)}
\nonumber\\
&= \hat{\rho}_{\rm ss} {\rm Tr}\hat{X}. 
\end{align}
This gives the third line in Eq.~(\ref{RK_a}). 
We note that Eq.~(\ref{RK_a}) leads to $\mathcal{K} \mathcal{R} \mathcal{K} = \mathcal{K}$. 
This implies that $\mathcal{R}$ satisfies one of the conditions for the Moore-Penrose pseudoinverse of $\mathcal{K}$.

\section{Derivation of Eq.~(\ref{vectorPotentialQME})}\label{appendix:vectorPotentialQME}

For completeness we here derive Eq.~(\ref{vectorPotentialQME}), 
the work version of the results in Refs.~\cite{SagawaHayakawa,Yuge_etal}. 
First we note that we can measure the work $W$ during varying the chemical potentials $\mu=\{\mu_i\}_i$ 
with a time interval $\tau$ as follows. 
At the initial time $t=t_0$, we perform a projection measurement of reservoir particle numbers $\{\hat{N}_i\}_i$ 
to obtain measurement outcomes $\{N_i(t_0)\}_i$. 
For $t>t_0$, we vary $\mu$, where the system evolves with interacting with the reservoirs. 
At $t=t_0+\tau$, we again perform a measurement of $\hat{N}_i$ to obtain outcomes $\{N_i(t_0+\tau)\}_i$.
The difference of the outcomes gives the work $W=\sum_i [\mu_i(t_0+\tau) N_i(t_0+\tau) - \mu_i(t_0) N_i(t_0)]$. 
Repeating the measurements, we obtain a probability distribution $p_\tau(W)$. 
The average work is given by $\langle W \rangle_\tau = \int dW p_\tau(W) W$, 
and the average work flow in a NESS is given by 
$J_W = \lim_{\tau\to\infty} \langle W \rangle_\tau / \tau$ with $\mu$ being fixed. 

In the following, we calculate the average work by 
$\langle W \rangle_\tau = \partial G_\tau(\chi) / \partial(i\chi)|_{\chi=0}$, 
where $G_\tau(\chi) \equiv \ln \int dW p_\tau(W) e^{i\chi W}$ is the cumulant generating function 
and $\chi$ is the counting field. 
By using the full counting statistics \cite{EspositoHarbolaMukamel_RMP}, we can calculate $G_\tau(\chi)$ by 
\begin{align}
G_\tau(\chi) = \ln {\rm Tr}_{\rm S} \hat{\rho}^{\chi}(\tau). 
\end{align}
Here $\hat{\rho}^\chi$ is the solution of the generalized quantum master equation (GQME): 
\begin{align}
\frac{\partial \hat{\rho}^{\chi}(t)}{\partial t} 
= \mathcal{K}^\chi\bigl(\mu(t)\bigr) \hat{\rho}^{\chi}(t), 
\label{GQME}
\end{align}
where the generalized generator is given by 
$\mathcal{K}^\chi = [ \hat{H}, ~] /i\hbar + \sum_j \mathcal{L}_j^\chi$, 
with the generalized dissipator $\mathcal{L}_j^\chi \hat{\rho} \equiv - (1/\hbar^2) \int_0^\infty dt' {\rm Tr}_j 
\bigl[ \hat{H}_{{\rm S}j} , [ \breve{H}_{{\rm S}j}(-t') , \hat{\rho} \otimes \hat{\rho}_j (\mu_j ) ]_\chi \bigr]_\chi$.
${\rm Tr}_j$ is the trace over the $j$th reservoir, 
$\hat{H}_{{\rm S}j}$ is the interaction Hamiltonian between the system and the $j$th reservoir, 
$\breve{H}_{{\rm S}j}$ is its interaction picture, 
$\hat{\rho}_i (\mu_j )$ is the thermal equilibrium state of the $j$th reservoir with the chemical potential $\mu_j$, 
$[\hat{O}_1 , \hat{O}_2]_\chi \equiv \hat{O}_1^\chi \hat{O}_2 - \hat{O}_2 \hat{O}_1^{-\chi}$, 
and $\hat{O}^\chi \equiv e^{-i\chi \sum_j \mu_j \hat{N}_j/2} \hat{O} e^{i\chi \sum_j \mu_j \hat{N}_j/2}$. 
Note that, if we set $\chi=0$, the GQME (\ref{GQME}) reduces to the original QME (\ref{QME}), 
and $\mathcal{K}^\chi$, $\hat{\ell}^\chi_0$, and $\hat{r}^\chi_0$ also reduce to $\mathcal{K}$, $\hat{1}$, 
and $\hat{\rho}_{\rm ss}$, respectively. 

For fixed $\mu$, we can define the left and right eigenvectors of  $\mathcal{K}^\chi(\mu)$ 
corresponding to the eigenvalue $\lambda^\chi_m(\mu)$, 
which are respectively denoted by $\hat{\ell}^\chi_m(\mu)$ and $\hat{r}^\chi_m(\mu)$.
They are normalized as ${\rm Tr} (\hat{\ell}^{\chi\dag}_m \hat{r}^\chi_n) = \delta_{mn}$.
We assign the label for the eigenvalue with the maximum real part to $m=0$.
It is known that $\lim_{\tau\to\infty} G_\tau(\chi)/\tau =\lambda^\chi_0$ holds \cite{EspositoHarbolaMukamel_RMP}.
Therefore, the average work flow $J_W$ in the NESS can be calculated by 
\begin{align}
J_W(\mu) = \frac{\partial \lambda^\chi_0(\mu)}{\partial (i\chi)}\bigg|_{\chi=0}.
\label{workFlux}
\end{align}

We now derive Eq.~(\ref{vectorPotentialQME}). 
We first note that the excess work can be written as 
$W_{\rm ex} = \partial G_{\rm ex}(\chi) / \partial(i\chi)|_{\chi=0}$, where 
$G_{\rm ex}(\chi) \equiv G_\tau(\chi) - \Lambda^\chi_0(\tau)$ and 
$\Lambda^\chi_m(t) \equiv \int_{t_0}^{t_0+t} dt' \lambda^\chi_m \bigl(\mu(t')\bigr)$. 
This is because $\langle W \rangle_\tau = \partial G_\tau(\chi) / \partial(i\chi)|_{\chi=0}$ and Eq.~(\ref{workFlux}).
To calculate $G_{\rm ex}(\chi)$, we solve the GQME (\ref{GQME}). 
For this purpose we expand $\hat{\rho}^\chi(t)$ as 
\begin{align}
\hat{\rho}^\chi(t) = \sum_m c_m(t) e^{\Lambda^\chi_m(t)} \hat{r}^\chi_m \bigl(\mu(t)\bigr).
\label{expansion_rho}
\end{align}
Substituting this expansion into Eq.~(\ref{GQME}) and taking the Hilbert-Schmidt inner product 
with $\hat{\ell}^\chi_0 \bigl(\mu(t)\bigr)$, we obtain 
\begin{align}
\frac{d{c}_0(t)}{dt} 
= - \sum_m c_m(t) e^{\Lambda^\chi_m(t)-\Lambda^\chi_0(t)} 
{\rm Tr}_{\rm S} \left[ \hat{\ell}^{\chi\dag}_0 \bigl(\mu(t)\bigr) 
\dot{\hat{r}}^\chi_m \bigl(\mu(t)\bigr) \right].
\nonumber
\end{align}
If the time scale of varying $\mu$ is sufficiently slower than the relaxation time of the system, 
we can approximate the sum on the RHS by the contribution only from the term with $m=0$ (adiabatic approximation).
By solving the approximate equation we obtain 
\begin{align}
c_0 (t_0+\tau) 
= c_0(t_0) \exp\left\{ - \int_{C} {\rm Tr}_{\rm S} \left[ \hat{\ell}^{\chi\dag}_0(\mu) 
d \hat{r}^\chi_0(\mu) \right] \right\}, 
\label{c_0}
\end{align}
where $C$ is a path connecting $\mu(t_0)$ and $\mu(t_0+\tau)$ in the parameter space and 
$d \hat{r}^\chi_0(\mu) \equiv \sum_j \bigl(\partial \hat{r}^\chi_0(\mu) / \partial \mu_j \bigr) d \mu_j$.
If $\hat{\rho}^\chi(t_0) = \hat{\rho}_{\rm ss} \bigl(\mu(t_0)\bigr)$, 
then $c_0(t_0) = {\rm Tr} \Bigl[ \hat{\ell}^{\chi\dag}_0\bigl(\mu(t_0)\bigr) 
\hat{\rho}_{\rm ss}\bigl(\mu(t_0)\bigr) \Bigr]$.

At long time, only the $m=0$ term remains in Eq.~(\ref{expansion_rho}) 
since $\Lambda^\chi_0(t)$ has the maximum real part. 
Therefore we obtain 
\begin{align}
\hat{\rho}^\chi(t_0+\tau) &\simeq c_0 (t_0+\tau) e^{\Lambda^\chi_0(\tau)} \hat{r}^\chi_0 \bigl(\mu(t_0+\tau)\bigr).
\end{align}
Substituting Eq.~(\ref{c_0}) into this equation we obtain an expression for 
$G_{\rm ex}(\chi) = \ln {\rm Tr}_{\rm S} \hat{\rho}^\chi(t_0+\tau) - \Lambda^\chi_0(\tau)$ as 
\begin{align}
G_{\rm ex}(\chi) = & - \int_C {\rm Tr}_{\rm S} \Bigl[ \hat{\ell}_0^{\chi\dag}(\mu) d \hat{r}_0^\chi(\mu) \Bigr] 
+ \ln {\rm Tr}_{\rm S} \Bigl[ \hat{\ell}_0^{\chi\dag} \bigl(\mu(t_0)\bigr) \hat{\rho}_{\rm ss} \bigl(\mu(t_0)\bigr) \Bigr] 
\nonumber\\
&+ \ln {\rm Tr}_{\rm S} \hat{r}_0^{\chi} \bigl(\mu(t_0+\tau)\bigr) . 
\label{CGF_ex}
\end{align}
Finally, by differentiating Eq.~(\ref{CGF_ex}) with respect to $i\chi$ and setting $\chi=0$, 
we obtain an expression for the excess work: 
\begin{align}
W_{\rm ex} 
&= - \int_C {\rm Tr}_{\rm S}\left[ \hat{\ell}_0^{\prime\dag}(\mu) d\hat{\rho}_{\rm ss}(\mu) \right]. 
\label{excess_derived}
\end{align}
We thus obtain Eq.~(\ref{vectorPotentialQME}).

\end{document}